\documentclass[twocolumn,showpacs,preprintnumbers,amsmath,amssymb,aps]{revtex4}
\usepackage{graphicx}
\begin{document}

\title{Electrophoresis of DNA on a disordered two-dimensional substrate}
\author{C.J. Olson Reichhardt and C. Reichhardt}
\affiliation{
Theoretical Division and Center for Nonlinear Studies,
Los Alamos National Laboratory, Los Alamos, New Mexico 87545}

\date{\today}
\begin{abstract}
We propose a new method for electrophoretic separation 
of DNA in which 
adsorbed polymers are driven over a disordered two-dimensional substrate
which contains attractive sites for the polymers.
Using simulations of a model for long polymer chains, we show that
the mobility {\it increases} with polymer length, in contrast
to gel electrophoresis techniques, and that separation can be achieved
for a range of length scales.  
We demonstrate that the separation mechanism relies on steric interactions
between polymer segments, which prevent substrate disorder sites from trapping
more than one DNA segment each.  Since thermal activation does not play a
significant role in determining the polymer mobility,
band broadening due to diffusion can be avoided in our
separation method.
\end{abstract}
\maketitle

\vskip2pc
\section{Introduction}
The development of new methods for efficiently separating charged biopolymers 
by length has been an area of significant recent activity
due to the fact that strategies for genome sequencing are based on sorting
DNA fragments by size \cite{Viovy}.
Simple charge-based sorting is not possible because 
the increase in electrostatic force on longer molecules with greater
total charge is exactly offset by a corresponding 
increase in hydrodynamic drag \cite{Slater86}. 
Instead, separation is achieved using techniques such
as gel electrophoresis, in which longer molecules are
slowed relative to shorter ones due to 
interactions with cross links in the gel.
Gel and capillary electrophoretic techniques are limited to
DNA strands less than $4\times 10^4$ base pairs (bp) in length
\cite{Schwinefus}; the mobility saturates for longer strands, and
sufficiently large strands fail to pass through the gel at all.
There is a need for separation of strands up to $1\times 10^6$ bp, 
and thus 
new techniques which can sort longer molecules are of particular interest
\cite{Huang96Ashton}.
The motion of elastic strings through random media is also of general
interest for a wide range of systems including 
magnetic domain wall motion, vortex lattice
motion in superconductors, and charge density waves.

Several recent proposals for electrophoretic techniques 
move away from the traditional media of gels and polymers
and instead take advantage of advances in nanolithography to
create microstructured devices for separation
\cite{Volkmuth,Dukeetc,Han2etc,Turner,Chou99,Han99,Rousseau97,Bakajin01,Huang02,Huang022,Cabodi02}.
The sorting effectiveness of these techniques is limited by
the relative size of the nanofabricated structure and the 
polymers to be sorted, making it necessary to fabricate 
a separate device for each size range of interest.
In contrast, Seo {\it et al.} \cite{Seo} proposed an
adsorption-based separation technique
that could permit the sorting of polymers which vary in size by three 
orders of magnitude.
When the ionic strength of the buffer solution is altered
\cite{Menes},
the DNA is partially adsorbed onto a clean surface, 
forming a series of loops which extend into the solution 
and trains which are adsorbed on the substrate.
It has been proposed that separation occurs because the longer 
polymers have a larger number of train segments, and thus 
experience a greater retardation of their motion \cite{Pernodet00,Luo}.  

The quasi-two dimensional geometry considered in Ref.~\cite{Seo} is
very appealing for separation purposes, in part because adsorbed
polymers spread out significantly into flat `pancakes'
\cite{deGennes} compared
to their coiled three-dimensional configurations, permitting better
coupling to length differences.
The sorting mechanism in Ref.~\cite{Seo} precludes complete
adsorption, however, since there is no separation for fully desorbed or fully
adsorbed polymers.  This limits the length range that can be processed,
since if the surface is strongly
attractive to DNA, long DNA chains fully adsorb and separation by length
is lost.  If instead the surface weakly attracts DNA, short chains
desorb from the surface and cannot be separated \cite{Seo04}.  

Here we propose an alternative sorting technique for long DNA strands
in which the polymers are fully adsorbed on the surface.
To permit separation, we spatially modify the surface, but instead of
of using posts or other impenetrable barriers, 
we consider randomly spaced pinning sites which temporarily retard the motion
of the polymer, yet still allow it to pass through.  
Such pinning could be created via 
the manipulation of
lipid bilayer membranes 
\cite{Hovis2etc}
or surface patterning \cite{Workman}.
We show that in this geometry, longer polymers are more mobile than
shorter ones, in contrast to typical
separation methods where longer polymers move
more slowly.  This 
avoids the jamming or clogging associated with long
polymers in other techniques.
The steric interaction between polymer segments causes the longer polymers
to be less well pinned by the random disorder
than the short polymers, and allows separation
by length to occur.

To demonstrate our separation mechanism, we use a simulation model 
that we have developed for long DNA fragments.
Many of the existing simulation models for electrophoretic processes
are best suited for shorter polymers
\cite{Olveraetc}.
Since we are concerned with polymers up to 300 $\mu$m in length, we do
not attempt to simulate each atom in the polymer.  Instead,
we adopt a bead-spring model in which the polymer is represented by 
multiple beads 
\cite{Rouse53} which are each spaced many persistence lengths apart.  
There is an entropic resistance
to the stretching of the polymer segment between two beads, which is
represented by a finitely extensible nonlinear spring (FENE) potential
\cite{Warner} that replaces the internal degrees of freedom of the 
polymer molecule \cite{Doyle97}.
An essential assumption of this model is that the polymer segment
between beads is significantly longer than the polymer persistence length.
This is in contrast to bead-stick models \cite{Kramer}, 
where the distance between beads is 
ten or less actual chemical segments.

\section{Simulation}
We employ Brownian dynamics \cite{Ermak}, 
permitting us to use time steps of order 0.1 ns,
orders of magnitude greater than the sub-fs time steps required in
all-atom molecular dynamics.  
In this technique, the solvent is treated statistically rather
than explicitly \cite{Rudisill92}.
The dimensionless 
force on bead $i$ in a chain $L$ 
base pairs long represented by $N$ beads is given by
\begin{equation}
{\bf F}_i=
\sum_{n.n.}{\bf F}_{i}^{FENE}
+\sum_{j=1}^N{\bf F}_{ij}^{EV}
+\sum_{k=i}^{N_p}{\bf F}_{ik}^{S}+{\bf F}^{E}
+{\bf F}^T,
\end{equation}
where ${\bf F}^{FENE}$ is the spring force along the chain,
${\bf F}^{EV}$ represents the excluded volume between beads,
${\bf F}^{S}$ is the force from a disordered substrate,
${\bf F}^{E}$ is the electrophoretic force, and ${\bf F}^{T}$ is
a thermal noise term.  
Distances are measured in terms of $\sigma_s$, 
the root mean square length of the spring.
Forces are expressed in terms of $k_BT/\sigma_s$.
We can neglect hydrodynamic interactions since
they are screened due to the proximity to the solid substrate
\cite{Maier,Bakajin98}.  
Electroosmotic effects can be controlled 
in the usual way by means of a high concentration
buffer \cite{Han99,Seo,Huang022,Han02}.  
We assume that 
the Debye length is considerably smaller than the distance between the beads
in our model.

The force between bead $i$ and neighboring beads is given by
\begin{equation}
F_{i}^{FENE}=\frac{-HQ}{1-(Q/Q_0)^2}\hat{\bf Q}
\end{equation}
where $Q=|{\bf l}-{\bf l}_0|$ 
is the elongation of the spring, ${\bf l}$ is the distance vector between
two neighboring beads, ${\bf l}_0=\sigma_s {\bf \hat{l}}$ is the equilibrium
spring length,
$Q_0$ is the maximum allowable 
elongation, and the Hookean spring constant 
$H=3/\sigma_s^3$. 
This phenomenological spring potential 
\cite{Warner} has the properties that it is
equivalent to a Hookean spring for small $Q$, but becomes infinite at
finite spring elongation.
The persistence length $l_p$ of double-stranded DNA is $l_p \approx 500$ \AA
\cite{Reese}.  The
Kuhn length $b_k=2l_p$, giving $b_k=0.1 \mu$m, where we assume that
the ionic strength of the buffer solution is sufficient to screen
electrostatic repulsion between sections of the chain \cite{Patel}.
Each base pair is 0.34 nm long so one Kuhn length contains 300 bp \cite{Long}.
Since we will be using a Gaussian chain model for the excluded volume
interactions, which requires the chain segments between beads to be
represented statistically and not deterministically,
the number $n$ of Kuhn lengths between beads must be sufficiently
large (well above the bead-rod limit of $n=1$), 
so we take $Q_0=nb_k=1.6\mu$m.  This gives $n=16$ and 
$\sigma_s=\sqrt{n}b_k=0.4\mu$m. 

The excluded volume term accounts for the repulsive interaction between
polymer segments when they approach each other \cite{Akkermans01,Kumar03},
an effect which is more pronounced in two dimensions than in
three dimensions \cite{Slater95}.
The excluded volume interaction for beads $i$ and $j$ a 
unitless distance $r_{bb}$ apart
is taken after that used
in Ref.~\cite{Jendrejack}, which is based on the energy penalty
due to overlap of two Gaussian coils, and has the form
\begin{equation}
{\bf F}_{ij}^{EV}=-Ar_{bb} e^{-B r_{bb}^2}\hat{\bf r}_{bb}
\end{equation}
where
\begin{equation}
A=\left(\frac{3}{4S_s^2}\right)^{5/2}\sigma_s v n^2\pi^{-3/2}, \ 
B=\frac{3\sigma_s^2}{4S_s^2}.
\end{equation}
Here the size parameter $S_s^2=nb_k^2/6$ \cite{Jendrejack}, while
the excluded volume parameter $v$ is taken to be $v=b_k^3$.
This gives
\begin{equation}
A=\frac{243\sqrt{2n}}{4\pi^{3/2}}, \  B=\frac{9}{2}.
\end{equation}

The substrate roughness is represented by 
$N_p$ finite range parabolic pinning traps
of radius $\sigma_p=0.4\sigma_s$, strength $f_p$, and density $\rho_p$.
The pin size was chosen to be close to the Kuhn length.
The force on bead $i$ from pin $k$ a distance $r_{bp}$ away is given by
\begin{equation}
{\bf F}_{ik}^{S}=f_p\frac{r_{bp}}{\sigma_p}\Theta(\sigma_p-r_{bp})\hat{\bf r}_{bp}
\end{equation}
where $\Theta$ is the Heaviside step function.

The electrophoretic force on each bead from an applied electric field
$E$ is
\begin{equation}
{\bf F}^{E}=qE\hat{\bf y}
\end{equation}
Here, $q=\lambda nb_k$ is the charge per bead, where 
the charge per unit length of DNA in solution is $\lambda=4.6\times 10^{-10}$
C/m, or $0.3 e^-/$\AA
\cite{Volkmuth}.
${\bf F}^{T}$ is the Langevin thermal noise term
representing the Brownian forces. 
It is a delta-correlated white noise process
which obeys
$\langle F^T\rangle=0$ and the fluctuation-dissipation theorem \cite{Kubo66}
$\langle F^T_i(t)F^T_j(t+\Delta\tau)\rangle =2k_BT\zeta^{-1}\delta_{ij}
\delta(\Delta \tau)$.
Here $\delta_{ij}$ is a Kronecker delta tensor and $\delta(\Delta \tau)$ is
the Dirac delta function.
Time is measured in units of $\tau=\zeta\sigma_s^2/(k_BT)$
and we take $\delta\tau=0.001$.
$\zeta$
is the friction coefficient characterizing the viscous 
interaction between the bead and the solvent.  We use the experimentally
measured value of $\zeta$ for polymers diffusing in a bilayer, 
$\zeta=2.97\times 10^{-7}$ Ns/m \cite{Olson}.
Theoretically, for a polymer moving in three dimensions,
$\zeta=6\pi\eta_s\sigma$, where
$\sigma$ is the effective bead radius and $\eta_s$ 
is the solvent viscosity.
We note that a theoretical expression for the friction coefficient in the case 
of a particle confined to a membrane suspended in a solvent has been developed
by Saffman \cite{Saffman75}, where only a weak dependence of $\zeta$ on
effective bead radius is obtained.

\begin{figure}
\includegraphics[width=3.5in]{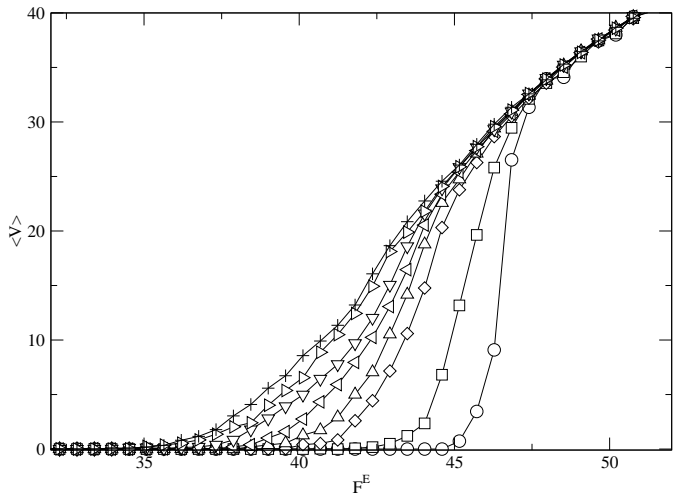}
\caption{
Average velocity $\langle V\rangle$ 
versus electric field for pinning density $\rho_p=0.8$,
pin strength $f_p=60$, and polymers of different length
represented by the number of beads
$N=5$ (circle), 10 (square), 20 (diamond), 30 (triangle up),
40 (triangle left), 50 (triangle right), 80 (triangle down), 
and 100 (plus).  
}
\label{figiv}
\end{figure}

\section{Results}
We first consider the velocity of the polymers over the rough 
substrate as a function of polymer length $L$.  We sweep the electric
field strength and find the average velocity $\langle V\rangle$ 
at each field value
during 200 repetitions of the sweep.
In Fig.~\ref{figiv} we plot the velocity-force curves for polymers of length
ranging from $N=5$ to 100, where $N$ is the number of beads used to
model the polymer, in a sample with
pinning density $\rho_p=0.8$ and strength $f_p=60$
at room temperature.  In physical units, this length range 
is $8\mu$m to 160$\mu$m, and it
includes $\lambda$-phage DNA, which has a contour
length of $21.2 \mu$m \cite{Hsieh}.
A velocity of 10 corresponds to
0.3 $\mu$m/s, and an applied field of 30 corresponds to 4.16 V/cm.
After the polymers depin, there is a range of driving force over which
we find a nonlinear velocity-force characteristic.
Within this range the shorter polymers
move more slowly than the longer polymers for a given electric field
strength.

\begin{figure}
\includegraphics[width=3.5in]{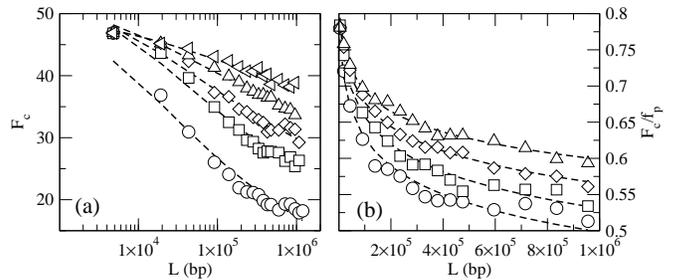}
\caption{
(a) Critical depinning force $F_c$
versus polymer length $L$ on a linear-log scale
for fixed $f_p=60$ 
and pinning density $\rho_p=$ 0.2 (circles), 0.4 (squares), 0.6 (diamonds),
0.8 (triangle up), and 1.0 (triangle left).  Dashed lines are logarithmic
fits.  (b) Scaled critical depinning force $F_c/f_p$ versus $L$
for fixed $\rho_p=0.8$ and $f_p=$ 20 (circles), 40 (squares), 60 (diamonds),
and 80 (triangles).  Dashed lines are logarithmic fits.
}
\label{figfc}
\end{figure}

Short polymers are better pinned by the underlying disorder than long
polymers, and thus a higher driving force must be applied before the
short polymers begin to move over the substrate.  
In Fig.~\ref{figfc} we illustrate
the dependence of the critical depinning force $F_c$ on polymer length
$L$ for a range of pinning strengths and densities.  Here we define
$F_c$ as the driving force at which $\langle V\rangle=1$.
We have chosen this definition since in our technique, physical separation
of the polymers will occur when the polymers are moving with different
velocities.  In an experiment, if the electric field were held at a
value between $F_c$ for polymers of two lengths, the two polymers will be
separated since one is strongly mobile while the other is nearly
immobile.
As shown in Fig.~\ref{figfc},
in each case we find that $F_c$ drops logarithmically with $L$, as 
indicated by the dashed lines.  As the pinning density is reduced
from $\rho=1.0$ to $\rho=0.2$, shown
in Fig.~\ref{figfc}(a), 
the depinning force drops and the variation of $F_c$ with
$L$ becomes steeper, meaning that the separation resolution is enhanced.
At the same time, the length range over which effective separation can
be achieved drops, and thus there is a trade-off which must be considered
depending on the range of sizes that are to be separated.  We show
the scaled length dependence of $F_c/f_p$ in Fig.~\ref{figfc}(b) for pinning
strengths $f_p=20$ to 80.  The pinning effectiveness drops slightly
faster than the pinning strength, as indicated by the fact that the curves
do not fall on top of each other.  For the weakest pins, the separation
effectiveness washes out above a length of $N\sim 100$.

\begin{figure}
\includegraphics[width=3.5in]{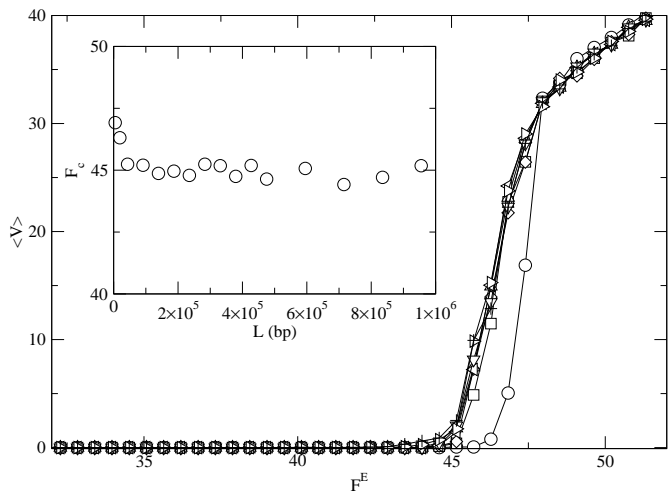}
\caption{
Average velocity $\langle V\rangle$ versus electric field for 
a system with no excluded volume at pinning density $\rho_p=0.8$,
pin strength $f_p=60$, and polymers of different length
$N=5$ (circle), 10 (square), 20 (diamond), 30 (triangle up),
40 (triangle left), 50 (triangle right), 80 (triangle down), 
and 100 (plus).  
Inset: $F_c$ versus $L$ for the same system.
}
\label{fignoint}
\end{figure}

The excluded volume
interactions play the key role in the separation mechanism.
What is happening physically can be
understood as follows.  Consider a polymer composed of only a single bead.  
This polymer can be completely trapped by a single pinning site.  Next, 
consider a polymer composed of two beads.  Although it is possible for the 
polymer to find two adjacent pinning sites such that both beads are pinned, 
it is more likely that one bead will be pinned while the other is still free.  
In this case the force from only one pin will have to hold two beads still 
against the electrophoretic force.  If the excluded volume interactions are 
removed, both beads can fit inside the pin, doubling the effective pinning
force.  As the polymer becomes longer and is composed of more beads, the
relative fraction of the polymer that is pinned decreases, provided that
each pin can capture only one bead.  This results in the decreased threshold
for depinning and the increased mobility of the longer polymers relative
to the short ones.  The importance of the excluded volume interaction is that
it enforces a pin occupancy of at most one bead per pin. 
Thus, the excluded volume interaction is what produces 
the decrease in $F_c$ with polymer length.
We test this by running a
series of simulations without the excluded volume interaction.  The
depinning force $F_c$ for this case is shown in the inset to 
Fig.~\ref{fignoint}, 
where it is clear that $F_c$ has no significant dependence on $L$.
The corresponding velocity-force curves are shown in Fig.~\ref{fignoint},
where it can clearly be seen that the curves lie on top of each other except
for the very shortest polymers.

\begin{figure}
\includegraphics[width=3.5in]{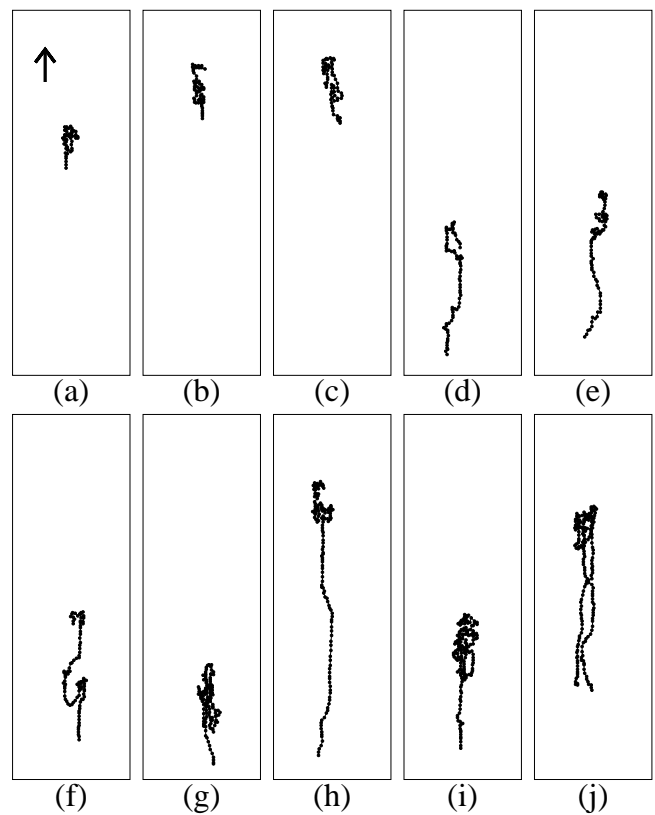}
\caption{
Images from polymer simulation showing bead positions for length $N=$
(a) 40, (b) 50, (c) 60, (d) 70, (e) 80, (f) 90, (g) 100, (h) 125,
(i) 150, (j) 175, and (k) 200.  The driving force is in the $+y$
direction, toward the top of the figure, as indicated by the arrow
in (a).
}
\label{figimage}
\end{figure}

We stress that the separation mechanism at work here is significantly
different than that which occurs in the case of impenetrable obstacles
such as cross links in gels or nanofabricated posts.  This can be seen
by observing 
images of the moving polymers.  A representative set of
images for polymers of different length is shown in Fig.~\ref{figimage}.
The chains are moving toward the top of the figure in the $+y$ direction.
Rather than forming hairpin structures, the polymers frequently form
a bundle on their advancing end, and sometimes drag one or two tail
segments.  

\begin{figure}
\includegraphics[width=3.5in]{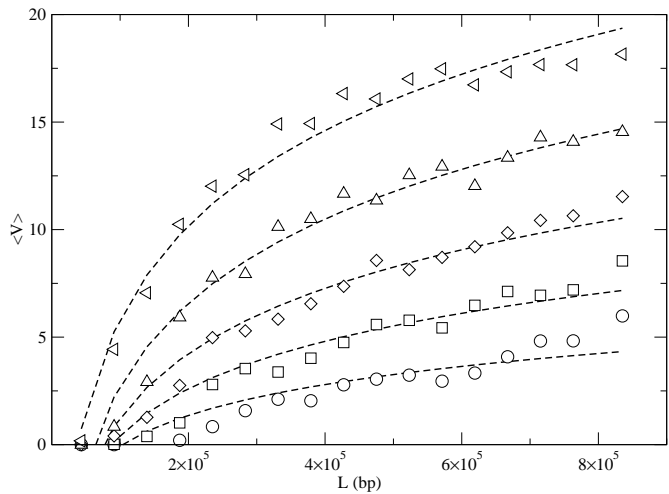}
\caption{
(a) Velocity $\langle V\rangle$
versus polymer length $L$ for $f_p=60$ 
and $\rho_p=0.8$ at different applied driving fields of
$F^E=$ 37.88 (circles), 39.0 (squares), 40.12 (diamonds),
41.24 (triangle up), and 42.36 (triangle left).  Dashed lines are logarithmic
fits.  
}
\label{figvel}
\end{figure}

For separation purposes, the polymer velocity must depend on length.
This can be achieved if the depinning force of the polymers is length
dependent.  To demonstrate this explicitly, we run a series of simulations
in which the driving force is held at a fixed value, and measure the
average velocity $\langle V\rangle$.  
The results are plotted in Fig.~\ref{figvel} for five
different values of $F^E$ in a sample with $\rho_p=0.8$ and $f_p=60$.
The velocity increases logarithmically with polymer length, as indicated by
the dashed lines, and velocity variations of an order of magnitude can
be achieved.

\begin{figure}
\includegraphics[width=3.5in]{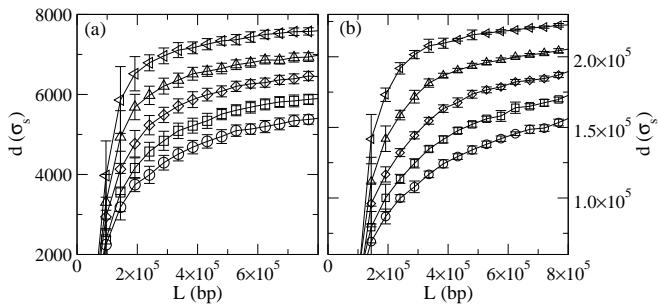}
\caption{
(a) Distance $d$ in units of $\sigma_s$ 
traveled over a fixed time interval 
versus polymer length $L$ for $f_p=60$ 
and $\rho_p=0.8$ at 
$F^E=$ 37.88 (circles), 39.0 (squares), 40.12 (diamonds),
41.24 (triangle up), and 42.36 (triangle left).  
Error bars indicate the
spread in distance traveled over 100 realizations of disorder.
(b) Distance $d$ traveled over a longer fixed time interval 
versus polymer length for the same system as in panel (a) with
$F^E=$ 37.88 (circles), 39.0 (squares), 40.12 (diamonds),
41.24 (triangle up), and 42.36 (triangle left).  
Error bars indicate the spread in distance traveled over 20 realizations
of disorder.
}
\label{figdistance}
\end{figure}

We note that in traditional gel electrophoresis techniques, 
diffusion is an important limiting effect, since the polymers are moving 
through the gel relatively slowly and depend on thermal fluctuations to help 
them translocate through the gel.  This effect is particularly pronounced for 
long polymers, which have extreme difficulty passing through the gel at all.
In contrast, in the technique proposed here, the polymers are much more
mobile than they would be in a gel.  The configurations and depinning of the
polymers are dominated by the strong electric fields and pinning imposed, and 
thermal effects play essentially no role in the separation.  We observe no 
significant thermal diffusion in our system at all.  As a result, diffusive 
broadening of the bands can be prevented.  The bands do still broaden due
to the intrinsic randomness of the pinning, which causes the progress of
the polymers over the substrate to be somewhat variable.  To illustrate
the magnitude of this broadening, in Fig.~\ref{figdistance}(a) we plot the
total distance traveled by the polymers under different drives applied for
a fixed period of time for 100 realizations of disorder.  Error bars indicate 
the average maximum and minimum distances traveled by polymers of a 
particular length.  
Higher resolution can be obtained by allowing the polymers to move a
larger distance through the gel, as illustrated in Fig.~\ref{figdistance}(b).

\begin{figure}
\includegraphics[width=3.5in]{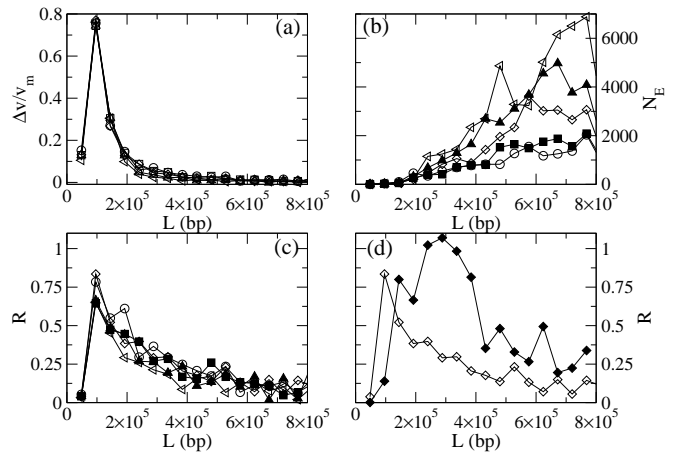}
\caption{
(a) Selectivity $\Delta V/V_m$ from the system in Fig.~\ref{figdistance}(a)
versus polymer length $L$ for $f_p=60$ 
and $\rho_p=0.8$ at 
$F^E=$ 37.88 (circles), 39.0 (squares), 40.12 (diamonds),
41.24 (triangle up), and 42.36 (triangle left).  
(b) Efficiency $N_E$ for the same system at
$F^E=$ 37.88 (open circles), 39.0 (filled squares), 40.12 (diamonds),
41.24 (filled triangle up), and 42.36 (triangle left).  
(c) Resolution $R$ for the same system at
$F^E=$ 37.88 (open circles), 39.0 (filled squares), 40.12 (diamonds),
41.24 (filled triangle up), and 42.36 (triangle left).  
(d) Resolution at $F^E=$ 40.12 for: (open diamonds) 
the system in Fig.~\ref{figdistance}(a) and
(filled diamonds) the system in Fig.~\ref{figdistance}(b) where the polymers
were allowed to travel a longer distance.
}
\label{figresolution}
\end{figure}

We quantify the separation power of our technique by measuring the
resolution as a function of polymer length.  The resolution is affected
by both the selectivity and efficiency of the separation for a given
length difference \cite{Petersen}.  The selectivity $\Delta V/V_m$ 
is proportional to the difference
in mobility for polymers of different length, 
\begin{equation}
\frac{\Delta V_i}{V_m}=\frac{2(\langle V_{i+1}\rangle - 
\langle V_{i}\rangle )}{\langle V_{i+1}\rangle +\langle V_i\rangle}
\end{equation}
where $\langle V_i \rangle$ is the average velocity for a polymer of
length $L_i$.
Fig.~\ref{figresolution}(a) shows that the selectivity 
for the same system in Fig.~\ref{figdistance}(a)
does not vary with
$F^E$ and is highest for the shortest polymers, in the same region
where Fig.~\ref{figvel} indicates that the
$\langle V \rangle$ versus $L$ curve has the steepest slope.
The efficiency $N_E$ is proportional to the width of the band 
$\sigma_ x^2$ observed after the polymers have traveled a distance $x$,
\begin{equation}
N_E=\frac{x^2}{\sigma_{x}^2} .
\end{equation}
As can be seen in Fig.~\ref{figresolution}(b), $N_E$ increases with both
$L$ and $F^E$, consistent with the decrease in the size of the error bars
at higher $L$ and $F^E$ shown in in Fig.~\ref{figdistance}(a).
The resolution $R$ is defined as
\begin{equation}
R=\frac{\sqrt{\bar N_E}}{4}\frac{\Delta V}{V_m}
\end{equation}
where $\bar N_E$ is the mean efficiency for the polymer lengths being
compared.  We plot $R$ versus $L$ in Fig.~\ref{figresolution}(c).  The
resolution depends more strongly on the selectivity than on the efficiency,
and as a result we find that $R$ is highest for the shortest polymers and
is not a strong function of $F^E$.  The resolution can be improved by
allowing the polymers to travel a longer distance, as in 
Fig.~\ref{figdistance}(b).  We compare the resolution for shorter and
longer distances traveled in Fig.~\ref{figresolution}(d), where we find
not only an enhancement of $R$ for the longer travel distance, but also a
shift in the peak value of $R$ towards longer polymers.  
This suggests that the technique 
could be optimized for separation of the desired range of $L$ by adjusting
the distance traveled by the polymers.

\section{Conclusion}
In summary, we have used a model developed for the simulation of long
DNA segments to demonstrate a new length separation mechanism
for polymers adsorbed to a disordered two-dimensional substrate.
Longer polymers are more mobile than short polymers, and the depinning
force decreases logarithmically with polymer length.  Correspondingly,
the polymer velocity increases logarithmically with length.  The 
separation mechanism arises due to the excluded volume interaction 
between chain segments, which serves to reduce the effectiveness of
the random pinning for longer polymers.
One possible experimental system
in which our proposed separation mechanism could be realized 
is solid-supported cationic lipid membranes,
where DNA is confined to two dimensions but free to diffuse
in plane \cite{Maier}.  
The pinning could be produced in the form of disorder on the supporting
substrate, which would perturb the bilayer and interfere with the free
diffusion of the DNA.  Such disorder could potentially be produced by
an experimental technique as simple as not fully cleaning the substrate 
before depositing the bilayer.
Our proposed separation mechanism offers several advantages over existing
techniques.  (1) It may not be necessary to use elaborate nanofabrication 
methods
to produce the pinning.  (2) The technique can be used to separate extremely 
long
strands of DNA which will not pass through conventional gels.
(3) It may be possible to achieve high throughput since the polymers do not
need to work their way around fixed impassible obstacles, but are instead
only temporarily hindered by the pinning sites, and can thus achieve much
higher overall mobilities than are possible in a gel, particularly for
long polymers.
(4) Since thermal effects do not play a significant role in the separation
technique, thermal broadening of the bands by diffusion should be strongly
suppressed.  The resolution limitation caused by band broadening from the
intrinsic disorder of the substrate can be reduced by allowing the polymers
to travel a longer distance during separation.

\section{Acknowledgments}
This work was supported by the U.S.~Department of Energy under
Contract No.~W-7405-ENG-36.

\end{document}